\documentclass[aps,prl,showpacs,twocolumn,superscriptaddress]{revtex4}
 \usepackage{amsmath}
\usepackage{graphicx}
\usepackage{epsfig}
\newcommand{\beq}{\begin{equation}}
\newcommand{\eeq}{\end{equation}}
\newcommand{\bea}{\begin{eqnarray}}
\newcommand{\eea}{\end{eqnarray}}

\begin{document}
\bibliographystyle{apsrev}
 \title{ Pseudospin entanglement and  Bell test in graphene }
\author{  M. Kindermann}
\affiliation{ School of Physics, Georgia Institute of Technology, Atlanta, Georgia 30332, USA  }

\date{July 2008 }
\begin{abstract} 
We propose a way of producing and detecting   pseudospin entanglement between electrons and holes in graphene. Electron-hole pairs are produced by a fluctuating potential and their entanglement is demonstrated by a current correlation measurement. The chirality of electrons   in graphene  facilitates a  well-controlled  Bell test with (pseudo-)spin projection angles  defined in {\em real} space. 
\end{abstract}
\maketitle

The entanglement between internal degrees of freedom of an electron and a hole in the Fermi sea is of both, fundamental and practical interest. It has been recognized as a form  of entanglement  that does not require many-body interactions \cite{beenakker:prl03} and comparatively simple ways of generating it experimentally have been proposed \cite{beenakker:prl03,samuelsson:prl04}. Yet, the   experimental demonstration of electron-hole entanglement in solid-state structures is still outstanding.  

Very recently Neder {\em et al.\ } \cite{neder:nat07} have    implemented     a quantum Hall interferometer that had been proposed \cite{samuelsson:prl04} for the generation and detection  of electron-hole entanglement.  This interferometer has allowed the observation of  the ``two-particle interference'' that is at the core of electron-hole entanglement \cite{neder:nat07}.  While this   is a first indication of entanglement production in the experiment  of Neder {\em et al.}, a conclusive verification of that entanglement has not yet been achieved.   It has been proposed  \cite{kawabata:jps01,chtchelkatchev:prb02,beenakker:prl03,samuelsson:prl04} that the generated entanglement is best verified through the violation of a Bell inequality. Experimentally, the demonstration of such a violation, however, meets with significant challenges. First,  the  decoherence in the interferometer of Ref.\ \cite{neder:nat07} needs  to be reduced significantly in order to safely preserve the  entanglement from the time of its production to its detection. In addition, a test of Bell inequalities requires measurements of spin-1/2 degrees of freedom along variable quantization axes. In the  setup of Ref.\ \cite{neder:nat07} these  quantization axes are defined by  scattering amplitudes that are poorly controlled experimentally, requiring an implementation  through trial and error.

In this Letter we develop a way of generating and detecting electron-hole entanglement that does not suffer from the above mentioned problems. We propose to entangle the  pseudospin \cite{novoselov:nma07} degree of freedom of electrons and holes in graphene \cite{novoselov:sci04,zhang:nat05,berger:jpc04} by means of the ``pumping'' mechanism of Refs.\ \cite{samuelsson:prb05,beenakker:prl05}.  The typical energy scales in graphene are considerably higher than those in GaAs, which has for instance allowed an observation of the quantum Hall effect at room temperature \cite{novoselov:sci07}. At comparable temperatures one thus expects  the decoherence   in graphene to be  much weaker than   in the experiment of Ref.\ \cite{neder:nat07}, addressing the first of the above  issues.   In addition, we formulate a Bell test through current correlation measurements  that overcomes  the mentioned problems of previously pursued   entanglement detection schemes \cite{kawabata:jps01,chtchelkatchev:prb02,beenakker:prl03,samuelsson:prl04}. Electrons in graphene have a definite chirality (for a  certain bandstructure ``valley''), moving in the direction of their pseudospin.   The pseudospin of an excitation can thus be measured through  its direction of motion. This affords a Bell test with straightforward and transparent control of the (pseudo)spin quantization axes that are now defined in {\it real} space. The   Bell test  proposed in Refs.\ \cite{kawabata:jps01,chtchelkatchev:prb02,beenakker:prl03,samuelsson:prl04}  is only valid in the regime of     temperatures $T$ that are low compared to the voltage $V$ applied to the interferometer: $  kT\ll eV$. Its application at finite temperatures  faces a  problem that has been discussed recently in Ref.\  \cite{hannes:prb08}.   The authors of Ref.\ \cite{hannes:prb08} suggest a cure of that issue whose  experimental implementation, however,   is  challenging: It  requires the addition of resonant levels to the setup.
Here we avoid the  problem   pointed out in Ref.\ \cite{hannes:prb08}  by a suitable postselection of the  entangled electron-hole pairs. That selection is implemented simply by  subtracting the  thermal background from all measured current correlators.
\begin{figure}\vspace{.5cm}
\includegraphics[width=8.5cm]{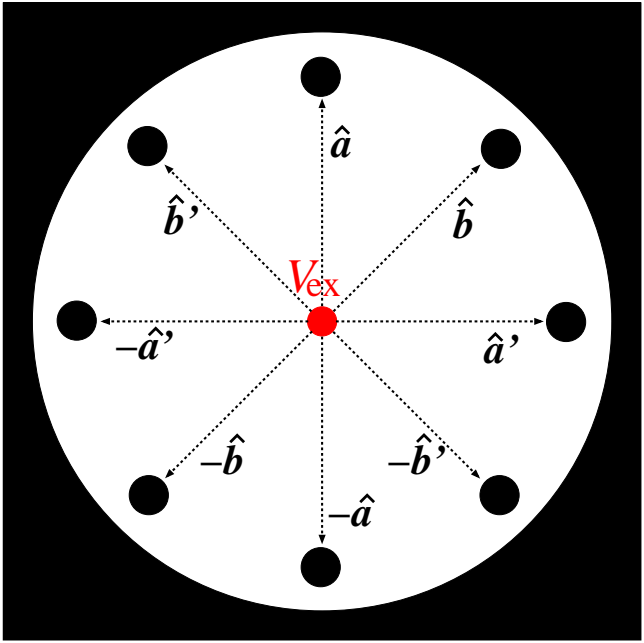}
\caption{ Proposed setup: A localized  fluctuating electric potential $V_{\rm ex}$ produces entangled electron-hole pairs in the center of a graphene sheet. The excitations are either drained at the rim of the sheet or leave into tunnel contacts at locations $r_\alpha\hat{\boldsymbol{\alpha}} $ (where $\alpha$ takes the values $\pm a$, $\pm a'$, $\pm b$, or $\pm b'$). The generated entanglement has signatures in correlations between the currents into these tunnel contacts. The starlike setup that is shown allows the maximal violation of a Bell inequality.}  \label{fig1} 
\end{figure}

{\em Setup:} We consider a sheet of ballistic graphene at low temperature $T$ and nonzero Fermi energy $\varepsilon_{\rm F}\gg kT$  in a  vanishing magnetic field $B$. The sheet is  well-coupled to an electron reservoir along its rim, as shown in Fig.\ \ref{fig1}. We formulate the low-energy Hamiltonian of graphene in   single-valley form  through  a unitary transformation that renders the Dirac model    valley-isotropic \cite{akhmerov:prl07},\beq \label{Dirac}
H_0= v\,  \boldsymbol{\sigma} \cdot \boldsymbol{p}  .
\eeq
Here, $\boldsymbol{ \sigma}=(\sigma_x, \sigma_y)$ is a vector of Pauli matrices in pseudospin space, $\boldsymbol{p}$  is the electron momentum, and $v$ is the Fermi velocity.  The graphene sheet is subject to a local fluctuating potential, as described by the Hamiltonian
\beq
H_{\rm ex}(t) = \int{ d\boldsymbol{x}   \, u(\boldsymbol{x}) \, eV_{\rm ex}(t)\,  \vec{\psi}^\dag(\boldsymbol{x}) \cdot  \vec{\psi}(\boldsymbol{x})}.
\eeq
 Here we have switched to second-quantized notation with electron annihilation operators $\vec{\psi}$ that are vectors in pseudospin space. The fluctuating potential is focused on a small region of  spatial extent $l_{\rm ex}$ in the middle of the graphene sheet and it is centered around the origin of our coordinate system. Its shape is given by a function $u$ that is normalized  to $ k^2_{\rm F}\int d\boldsymbol{x}\, u(\boldsymbol{x})=1$ [for instance   $u(\boldsymbol{x}) =  \exp(-|\boldsymbol{x}|^2/2l_{\rm ex}^2)/2\pi k^2_{\rm F} l^2_{\rm ex}$]. The length $l_{\rm ex}$  is assumed to be large compared to the lattice spacing $l_{\rm lattice}$, but small compared to the Fermi wavelength $2\pi/k_{\rm F}$, that is  $l_{\rm lattice} \ll l_{\rm ex} \ll 2\pi/k_{\rm F}$. We assume that the frequency spectrum of the potential correlator  $c_V(-\omega)=\int dt \,\exp(-i \omega t) \langle V_{\rm ex}(t)V_{\rm ex}(0)\rangle$   is relatively flat  until it vanishes above a high-frequency cut-off $\omega \approx \Omega$ with $kT\ll \Omega\ll \varepsilon_{\rm F}$ \cite{samuelsson:prb05,beenakker:prl05,footnote0}.
The fluctuating potential $V_{\rm ex}$ creates pairs of electrons and holes (in the sense of an electron missing in an otherwise filled Fermi sea) that propagate outward before they are reflectionlessly drained by the reservoir surrounding the graphene sheet. On their way to the rim they are able to leave through tunnel contacts $\alpha$  with sizes $l_{\alpha}\ll  {r}_\alpha $ into additional electron reservoirs   at  locations $\boldsymbol{x}= r_\alpha\hat{\boldsymbol{\alpha}} $, where $|\hat{\boldsymbol{\alpha}}|=1$ and $ {r}_\alpha  \gg 2\pi/k_{\rm F}$ (see Fig.\ \ref{fig1}),
\beq
H_{{\rm T},\alpha}= \int d\boldsymbol{x} \, \vec{\psi}^\dag(\boldsymbol{x} )\cdot \left(\begin{array}{cc} w^{\rm A}_\alpha (\boldsymbol{x} ) \\ w^{\rm B}_\alpha (\boldsymbol{x} )\end{array}\right) \psi^{\rm res}_\alpha +h.c.
\eeq
 \cite{footnote01}.
Here, the functions $\vec{w}_\alpha $  are centered around $\boldsymbol{x}= r_\alpha\hat{\boldsymbol{\alpha}} $  and the operator $\psi^{\rm res}_\alpha $ annihilates electrons in the reservoir of contact $\alpha$. All electron reservoirs    are in thermal equilibrium with the graphene sheet. Every tunnel contact  $\alpha$  has one counterpart $-\alpha$   in direction $-\hat{\boldsymbol{\alpha}} $.

 {\em Entanglement production:} We first consider the excitations created by a short potential pulse, $eV_{\rm ex}(t)=\zeta \delta(t-t_{\rm ex})$. In  first-quantized form the low energy contribution ($|\boldsymbol{p}-\boldsymbol{p}'| \ll k_{\rm F}$) to the electron-hole pair  that is produced at first order in $\zeta$ reads (we set $\hbar = 1$)
\beq  \label{Bell}
 |\psi(t)\rangle_\zeta\Big|_{t=t_{\rm ex}+0^+} = \zeta\sum_{\boldsymbol{pp}'} |\boldsymbol{p}\rangle^{\rm el } |\boldsymbol{p'}\rangle^{\rm h } \left( |\!\uparrow \rangle^{\rm el } |\!\uparrow\rangle^{\rm h }+ |\!\downarrow \rangle^{\rm el } |\!\downarrow\rangle^{\rm h }\right),
\eeq
where $\uparrow$ and $\downarrow$ specify the pseudospin direction and $|\rangle^{\rm el }$ and $|\rangle^{\rm h }$   are electron and  hole amplitudes, respectively (with the convention that a hole has the same  pseudospin  as the electron that it replaces).  
The pseudospins of the   electron and the hole  described by $|\psi\rangle_\zeta$, Eq.\ (\ref{Bell}), are entangled. They form a so-called Bell pair.   A source with a periodically varying potential $V_{\rm ex}(t)$   serves as a steady supply of such Bell pairs.   Excitations  that appear at higher order in $ \zeta $   are negligible if $|eV_{\rm ex}(t)| \ll \varepsilon_{\rm F}$ for all $t$, which we assume henceforth.

\begin{figure}\vspace{.5cm}
\includegraphics[width=8cm]{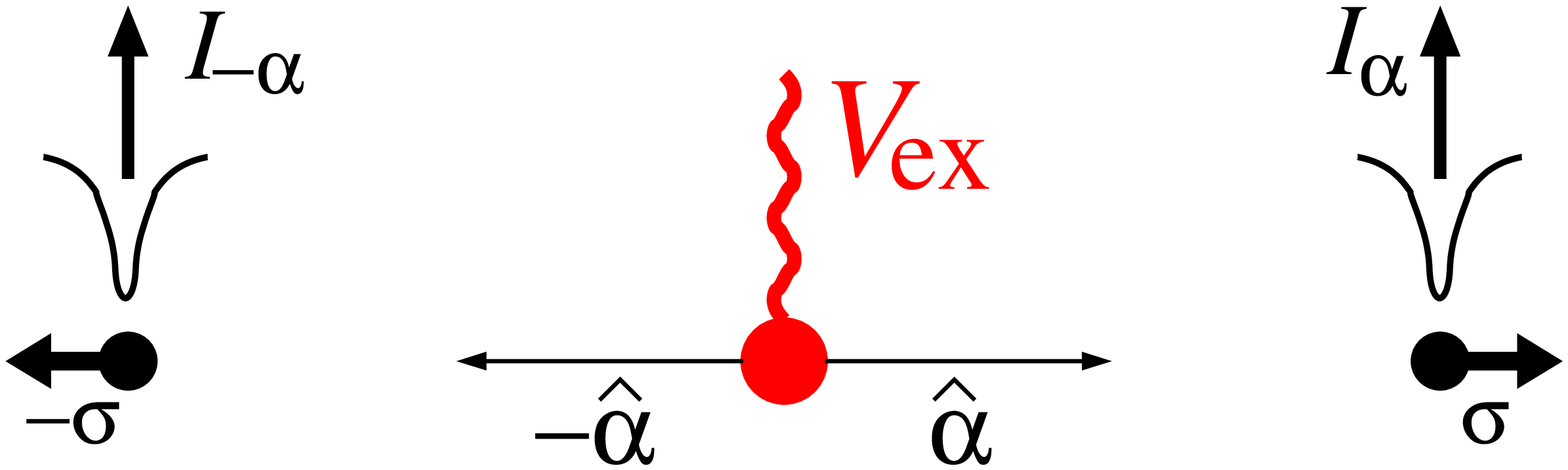}
\caption{Pseudospin measurement: Electron and hole excitations  are generated  by $V_{\rm ex}$ at $\boldsymbol{x}=0$. They contribute to the current $I_\alpha $ into contact $\alpha$  at $ \boldsymbol{x}=r_\alpha \hat{\boldsymbol{\alpha}}$ if  their velocity satisfies $\dot{\boldsymbol{x}}\,||\,\hat{\boldsymbol{\alpha}}$.  The pseudospin $\boldsymbol{\sigma} $ of these excitations is fixed due to their  chirality: one has $\boldsymbol{\sigma}\,||\,\dot{\boldsymbol{x}}$  and $\boldsymbol{\sigma} $ thus points ``up'' along the quantization axis $\hat{\boldsymbol{\alpha}}$.   The opposing contact $-\alpha$   at $ \boldsymbol{x}=-r_\alpha \hat{\boldsymbol{\alpha}}$ collects excitations with pseudospin  ``down'' along the same axis. 
}\label{fig2} 
\end{figure}

{\em Entanglement detection:} The Heisenberg equations of motion corresponding to the Hamiltonian $H_0$, Eq.\  (\ref{Dirac}),
\beq \label{dynamics}
\dot{\boldsymbol{p} }=0 ,\;\; \;\;\dot{\boldsymbol{ \sigma} }= 2 v \boldsymbol{p} \times\boldsymbol{ \sigma} ,\;\;\;\;  \dot{\boldsymbol{x} }=v\boldsymbol{ \sigma} ,
\eeq
 show that the pseudospin of an electron is not conserved. We therefore   postselect orbital states of the form $|p\rangle_\alpha  =(|p \hat{\boldsymbol{\alpha}} \rangle+|-p \hat{\boldsymbol{\alpha}} \rangle)/\sqrt{2}$. The initial wavefunction $|\psi\rangle_\zeta$, Eq.\ (\ref{Bell}), factorizes into   an isotropic orbital part and a pseudospin part. It follows    that at $t=t_{\rm ex}$ also the pseudospin state of all postselected electron-hole pairs is given by the pseudospin part  of Eq.\ (\ref{Bell}) and entangled. Moreover, the (unnormalized) density matrix of a postselected electron  $\rho_{pp'}^{\alpha}=\langle p|_\alpha\, \rho | {p}'\rangle_\alpha$,   $\rho$ being the  density matrix of the electron before postselection, takes the form $\rho^{\alpha}_{pp'}(t)=\cos [2vp(t-t_{\rm ex})]\cos [2vp'(t-t_{\rm ex})] \rho^{\alpha}_{pp'}(t_{\rm ex})$.  After normalization $\rho^{\alpha}$ is time-independent. The pseudospins of the  postselected excitations are   thus conserved and so is their entanglement. We propose to verify that entanglement  by violation of a  Bell inequality. This requires a measurement of the postselected pseudospins with variable quantization axes. The tunnel contacts $\alpha$ serve that purpose. To see this we integrate Eqs.\ (\ref{dynamics})  to find
\beq \label{project}
 \boldsymbol{x} (t) = \boldsymbol{x}(t_{\rm ex})  +v (t-t_{\rm ex}) \frac{ \left[  \boldsymbol{p}(t_{\rm ex}) \cdot \boldsymbol{ \sigma}(t_{\rm ex}) \right]   \boldsymbol{p}(t_{\rm ex})}{|\boldsymbol{p}(t_{\rm ex})|^2} + {\cal O} \left(\frac{1}{|\boldsymbol{p}|}\right).
\eeq
Consider an electron (before postselection) that is produced   at  $t=t_{\rm ex}$ and $\langle\boldsymbol{x}(t_{\rm ex})\rangle=0$, such that semiclassically the first term in Eq.\ (\ref{project}) vanishes.  The third, oscillatory term in Eq.\ (\ref{project}) is smaller than the second one by a factor $(k_{\rm F} r_\alpha )^{-1}$ and negligible in our limit. Assume that this electron is detected in contact $ \alpha$     at time $t$, such that $\boldsymbol{x}(t)=  r_{\alpha} \hat{\boldsymbol{\alpha}}$. This projects the initial state of the electron onto  eigenstates of $\boldsymbol{x}(t)$, Eq.\ (\ref{project}), with eigenvalue $ r_{ \alpha} \hat{\boldsymbol{\alpha}}$.      To our accuracy, when only the second term in Eq.\ (\ref{project}) is relevant, such a measurement projects onto  amplitudes   with  $\boldsymbol{p}  \,|| \hat{\boldsymbol{\alpha}}$. 
Moreover,  substituting $ \boldsymbol{p}(t_{\rm ex})/|\boldsymbol{p}(t_{\rm ex})|= \hat{\boldsymbol{\alpha}}$  into Eq.\ (\ref{project}) (the sign of $\boldsymbol{p}(t_{\rm ex})$ is arbitrary since in enters quadratically), we see that   a measurement of an electron in contact $\alpha$  projects onto amplitudes with    $\boldsymbol{\sigma}(t_{\rm ex})  \cdot  \hat{\boldsymbol{\alpha}}= 1$.  Likewise, detection of an electron in contact $-\alpha$   projects onto amplitudes   with  $\boldsymbol{\sigma}(t_{\rm ex})   \cdot  \hat{\boldsymbol{\alpha}} = -1$ \cite{footnote2}. We conclude that   the tunnel   contacts   $ \alpha$  collect currents $I_\alpha$ of electrons and holes with a definite initial pseudospin when measured along   quantization axes $ \hat{\boldsymbol{\alpha}}$  that are defined by the locations of the   contacts $ \alpha$  in {\em real} space, as illustrated in Fig.\ \ref{fig2}. 
As discussed above,  the initial pseudospin state  of the created electron-hole pairs before postselection  equals the pseudospin state  of any of the postselected electron-hole pairs. Also the pseudospin of the postselected pseudospins may thus be inferred from a measurement of the currents $I_\alpha$.

In order to demonstrate the  entanglement of the postselected electron-hole pairs we formulate a slightly modified  Clauser-Horne-Shimony-Holt  inequality 
\beq \label{CHSH}
{\cal B} \leq 2 \; {\rm with}\;\; {\cal B}= C_{ab} + C_{a'b} +C_{ab'} -C_{a'b'} 
\eeq
  in terms of symmetrized correlators 
\beq \label{Cabdef}
C_{ ab}=\frac{1}{2}\langle \psi | (\hat{\boldsymbol{a}}\cdot \boldsymbol{\sigma}^{\rm el}  )(\hat{\boldsymbol{b}} \cdot \boldsymbol{\sigma}^{\rm h}   )+(\hat{\boldsymbol{b}}\cdot \boldsymbol{\sigma}^{\rm el}  )(\hat{\boldsymbol{a}} \cdot \boldsymbol{\sigma}^{\rm h}   ) |\psi \rangle
\eeq
of  an electron pseudospin $\boldsymbol{\sigma}^{\rm el}$ and a hole   pseudospin $\boldsymbol{\sigma}^{\rm h}$,      projected onto the unit vectors $\hat{\boldsymbol{a}}$ and $\hat{\boldsymbol{b}}$. Only entangled electron-hole pairs  can violate the inequality (\ref{CHSH}) and the parameter ${\cal B}$ is thus an ``entanglement witness'' \cite{terhal:pla00}.
 
At zero temperature   and in our  limit of dilute electron-hole pairs $ |eV_{\rm ex}|\ll \varepsilon_{\rm F} $  one shows along the lines of Refs.\ \cite{kawabata:jps01,chtchelkatchev:prb02} that the parameter $C_{ ab}$ can be expressed   through   zero-frequency current correlators. One needs to correlate the currents  of electrons and holes with definite   pseudospins when measured along the quantization axes  $\hat{\boldsymbol{a}}$ and $\hat{\boldsymbol{b}}$, respectively.
 The reasoning of the paragraph around Eq.\ (\ref{project}) allows us to relate these currents to the tunnel currents $I_{\pm a}$ and $I_{\pm b}$ into the contacts  $\alpha=\pm a$  and $\alpha=\pm b$. In case all tunnel contacts     couple  with the same strength  to the relevant electrons or holes one  concludes in this way that \cite{kawabata:jps01,chtchelkatchev:prb02} 
 \beq \label{Cab}
C_{ ab}= \frac{  \sum_{\sigma,\sigma'=\pm 1} \sigma\sigma' c_{\sigma{a},\sigma'{b}} }{    \sum_{\sigma,\sigma'=\pm 1}  c_{\sigma{ a},\sigma'{ b}} },
\eeq
  where  $\tilde{c}_{{ a},{ b}}=\int dt \,  \langle \delta I_{ {a}}(t) \delta I_{ {b}}( 0)\rangle$ has to be substituted for $c_{a,b}$. 
 Unlike in the proposals of Refs.\   \cite{beenakker:prl03,samuelsson:prl04}, here not every produced electron-hole pair is detected. To find $c_{a,b}$ in the general case, when the tunnel coupling strengths to different reservoirs $\alpha$    are not equal, one  therefore has to normalize the correlators $\tilde{c}_{a,b} $   by the detection probabilities    $W_{\alpha}$, 
\beq \label{cex}
c_{{ a},{ b}}=\left(W_{a}W_{ b}\right)^{-1} \int dt \, \left[\langle \delta I_{ { a}}(t) \delta I_{ { b}}\rangle-\langle \delta I_{ { a}}(t) \delta I_{ { b}} \rangle_{V_{\rm ex}=0} \right].
\eeq 
The probabilities $W_\alpha $ can be measured through the AC-response of the currents $I_{\alpha}$ to the excitation potential $V_{\rm ex}$ at frequencies  $\omega \ll {\rm min}\{v/l_{\alpha},\varepsilon_{\rm F}\}$,
\beq \label{W}
W_{\alpha} =\left|\frac{\varepsilon_{\rm F}}{\omega}\right| \left|\frac{I_{\alpha}(\omega)}{e^2   V_{\rm ex}(\omega)}\right|.
\eeq
 
Extra care has to be taken at finite temperature. We avoid the issue pointed out in Ref.\  \cite{hannes:prb08} here by an additional postselection of the electron-hole pairs for which Eq.\ (\ref{CHSH}) is evaluated (see Appendix A for the details). 
 This selection is made by the subtraction of the equilibrium current correlations  in Eq.\ (\ref{cex}). We show in Appendix A that after that subtraction a violation of the Bell inequality (\ref{CHSH}) is an unambiguous signature of entanglement  also  at finite temperature,   as long as $v/|  r_a\hat{\boldsymbol{a}} \pm{r}_{b}\hat{\boldsymbol{b}}  | \ll kT \ll \Omega$.

 {\em Predictions:} In our limit of small $\vec{w}_\alpha $ (that is tunneling contacts),  $v/|  r_a\hat{\boldsymbol{a}} \pm{r}_{b}\hat{\boldsymbol{b}}  |\ll kT\ll\Omega\ll \varepsilon_{\rm F}$,   $ |eV_{\rm ex}|\ll \varepsilon_{\rm F}$, $l_{\rm ex} \ll 2\pi/k_{\rm F}$, and  $ l_\alpha  \ll {r}_\alpha $ we find (see Appendix B)
 \beq \label{cres}
c_{{ a},{ b}}= \frac{e^4}{2\varepsilon_{\rm F}^2} (1+\hat{\boldsymbol{a}}\cdot \hat{\boldsymbol{b}})   \int_{-\infty}^{\varepsilon_{\rm F}}d\varepsilon \int_{\varepsilon_{\rm F}}^{\infty}d\varepsilon' \,c_V(\varepsilon-\varepsilon').
 \eeq
 We conclude that the pseudospin correlators measured through Eqs.\ (\ref{Cab}) and (\ref{cex}) take the form
 \beq \label{Cabres}
 C_{ab}= \hat{\boldsymbol{a}}\cdot \hat{\boldsymbol{b}},
 \eeq
 which is immediately shown to violate the modified CHSH inequality (\ref{CHSH}) for appropriate choices of the vectors $\hat{\boldsymbol{a}}$, $\hat{\boldsymbol{a}}'$, $\hat{\boldsymbol{b}}$, and $\hat{\boldsymbol{b}}'$. The maximal violation $ {\cal B}=2 \sqrt{2} $  for instance can be  achieved in the symmetric starlike setup shown in Fig.\ \ref{fig1}, where the vectors $\hat{\boldsymbol{a}}'$, $\hat{\boldsymbol{b}}$, $\hat{\boldsymbol{a}}$, and $\hat{\boldsymbol{b}}'$ are separated by successive $45^\circ$ angles.
  
 For Eq.\ (\ref{Cabres}) to hold  the created excitations must not change their pseudospin on the way to the tunnel contacts, for instance through decoherence. We thus assume excitation energies such that the inelastic mean free path $l_{\rm in}$ is long,  $l_{\rm in}\gg {r}_\alpha $. This condition is fulfilled at  $\Omega \ll {\rm min} \{ v_{\rm ph} k_{\rm F}, \varepsilon_{\rm F}/\sqrt{k_{\rm F}{r}_\alpha }\}$, where $v_{\rm ph}$ is the phonon velocity in graphene \cite{sarma:prb07,hwang:prb08}.  As anticipated, for suitable parameter values decoherence through inelastic processes in our proposal is    already suppressed at temperatures much higher than those in the experiment of Ref.\ \cite{neder:nat07}.   We need in addition that  also the  elastic mean free path $l_{\rm el}$ is long,     $l_{\rm el}  \gg {r}_\alpha $. Long elastic mean free paths have been found in suspended sheets of graphene \cite{bolotin:08}. We conclude that the setup depicted in Fig.\ \ref{fig1}  allows to generate and conclusively demonstrate electron-hole entanglement if  ${\rm min}\{v_{\rm ph} k_{\rm F},\varepsilon_{\rm F}/\sqrt{k_{\rm F} {r}_\alpha }\}\gg\Omega \gg kT \gg v/|  r_a\hat{\boldsymbol{a}} \pm{r}_{b}\hat{\boldsymbol{b}}  |$.
 
 {\it Discussion: } We have proposed a way of creating and  verifying pseudospin entanglement of electron-hole pairs in graphene. Bell pairs are   produced by a fluctuating potential and their entanglement is demonstrated after postselection through violation of a Bell inequality.  The quantization axes in the requisite pseudospin measurement  are defined by the locations of tunnel contacts 
  in {\em real} space.  This simplicity of the pseudospin measurement is bought at a price: the postselected Bell pairs are not easily  separated spatially, since  the advocated pseudospin measurement is nonlocal, as shown in Fig.\ \ref{fig2}. The produced Bell pairs, however,   are entangled also with  respect to   their intrinsic spins \cite{footnote21},   which entanglement is readily spatially separated  \cite{samuelsson:prb05,beenakker:prl05}. The proposed experiment is thus an intermediate step toward the generation and manipulation of spatially separated Bell pairs in electronic structures.  Entanglement is generated by a mechanism that is able to produce  spatially separated Bell pairs, but it is detected before that spatial separation is achieved \cite{footnote3}.    The proposed detection mechanism affords three major advantages:    i)   it suffers less from  decoherence than previously pursued implementations of particle-hole entanglement; ii) it allows a well-controlled Bell test with clearly defined (pseudo)spin quantization axes;  iii) it avoids  problems with earlier proposals of electron-hole entanglement detection at finite temperature.   Our proposal thus overcomes some critical hurdles on the way to an  observation of particle-hole entanglement in electronic structures. 
  
  The author thanks W.\ A.\ de Heer, P.\ N.\ First, and L.\ You very much for discussions.

 \vspace{2cm}
 
  \begin{appendix}
  
  \section{Appendix A: Bell test} \label{appA}
  The relation between current correlations and the (pseudo)spin correlator $C_{ab}$, Eq.\ (\ref{Cab}), has been  proven in Refs.\ \cite{kawabata:jps01,chtchelkatchev:prb02} at zero temperature in the limit  of dilute electron-hole pairs  (in the system discussed here that is $|eV_{\rm ex}| \ll \varepsilon_{\rm F}$).  It has been pointed out recently \cite{hannes:prb08} that  extra care has to be taken at finite temperature. Thermal excitations in the reservoirs that collect the measured currents are then able to generate electron and hole currents that flow in the ``wrong'' direction: from the reservoirs into the system that contains the entangled electron-hole pairs.  These currents pose a serious problem and they can invalidate a Bell test along the lines of   Refs.\ \cite{kawabata:jps01,chtchelkatchev:prb02}  at finite temperature. 
In Ref.\  \cite{hannes:prb08} an alternative, energy-selective detection scheme   has been proposed that does not suffer from the same problem. Experimental  implementations of energy-resolved detection, for instance through resonant levels \cite{kindermann:prl06}, are, however, expensive. In contrast to the situation studied in Ref.\  \cite{hannes:prb08} the entangled electrons and holes in our proposal  differ in energy. This affords a simpler cure of the problem.
  
 The typical energy separation between the electron and the hole in the produced Bell pairs  is $\Omega$. This allows us to restrict our attention  by means of  postselection to electron-hole pairs with electron  energies  $ \varepsilon >\varepsilon_{\rm F} + \omega$ and hole energies $\bar{\varepsilon} <\varepsilon_{\rm F} - \omega$, where we choose $kT\ll \omega \ll \Omega$.  Thermally activated electron-hole pairs in the graphene sheet and the reservoirs coupled to it have typical energies $|\varepsilon-\varepsilon_{\rm F}| , |\bar{\varepsilon}-\varepsilon_{\rm F}|\approx kT$.  All but an exponentially suppressed number of them are excluded by the above postselection. This avoids the problem that has been pointed out in Ref.\  \cite{hannes:prb08}.  At the same time our postselection includes almost all excitations created by $V_{\rm ex}$, that have typical energies $|\varepsilon-\varepsilon_{\rm F}|, |\bar{\varepsilon}-\varepsilon_{\rm F}|\approx \Omega$. The above postselection may thus   be implemented approximately by  subtracting the statistical (thermal) contributions from all measured current correlators, leaving only correlations due to $V_{\rm ex}$. These  thermal correlations, in turn, may be inferred from a measurement of  the respective correlators  in equilibrium, at $V_{\rm ex}=0$. The above postselection should thus solve the problem pointed out in Ref.\ \cite{hannes:prb08} with very moderate additional experimental effort: it is  implemented by a second measurement in thermal equilibrium, as expressed in  Eq.\ (\ref{cex}). Below we prove this expectation correct.

 The two-particle density matrix after our postselection of electron-hole pairs in the conduction band with symmetrized orbital states $|p\rangle^{\rm el}_\alpha$ and  $|\bar{p}\rangle^{\rm h}_{\bar{\alpha}}$ for electron and hole, respectively, at electron  momenta  $ {p} >(\varepsilon_{\rm F} + \omega)/v$ and hole momenta $\bar{p} <(\varepsilon_{\rm F} - \omega)/v$ reads
  \begin{widetext}
\bea \label{rho}
\rho^{  \,{\rm el-h}\,\alpha\bar{\alpha}}_{\sigma\bar{\sigma},\sigma'\bar{\sigma}'}( {p},  \bar{p} ;{p}', \bar{p}' )& \propto& \Theta(v {p}-\varepsilon_{\rm F}-\omega) \Theta(\varepsilon_{\rm F} -\omega-v \bar{p}) \Theta(v {p}'-\varepsilon_{\rm F}-\omega) \Theta(\varepsilon_{\rm F} -\omega-v \bar{p}')  \\
&&  \mbox{}\times \langle   {p}  |_{\alpha}^{{\rm el}}\, \langle \sigma|^{\rm el} \,\langle   \bar{p}   |_{\bar{\alpha}}^{{\rm h}}\,\langle \bar{\sigma}|^{\rm h}  \,\rho_{\rm cond}  |    {p}'  \rangle_{\alpha}^{{\rm el}}\,| \sigma'\rangle^{\rm el} \, |    \bar{p}'\rangle_{\bar{\alpha}}^{{\rm h}}\,| \bar{\sigma}'\rangle^{\rm h}\nonumber
\eea
[$\Theta(x)=1$ for $x>0$ and  $\Theta(x)=0$ otherwise], where    the pseudospins of the electron and the hole are denoted $\sigma$ and $\bar{\sigma}$, respectively. $\rho_{\rm cond}$ is the density matrix of all electron-hole pairs in the conduction band of the graphene sheet (before postselection). 
  Of all momentum-restricted electron-hole pairs we then select those whose electron and hole are found within a radial distance $|r-\bar{r}|\lesssim \Delta r $ of each other (in real space), as described by the reduced density matrix
 \beq \label{rhospin}
 {\rho}^{{\rm  spin} }_{\sigma\bar{\sigma},\sigma'\bar{\sigma}'} \propto \int_{0}^R dr d\bar{r} \,e^{-(r-\bar{r})^2/2\Delta r^2} \,\tilde{\rho}_{\sigma\bar{\sigma},\sigma'\bar{\sigma}'}^{{\rm el-h}\, \alpha\bar{\alpha}} ( {r},  \bar{r};{r},\bar{r} ) .
 \eeq
  \end{widetext}
Here, $\tilde{\rho}^{\rm el-h}$ is the Fourier transform of $\rho^{\rm el-h}$ [with the sign convention of Eq.\ (\ref{psi})]  and $R$ is the radius of the graphene sheet. As explained in the main text, ${\rho}^{{\rm  el-h}\, \alpha\bar{\alpha}}$ is independent of $\hat{\boldsymbol{\alpha}}$ and $\hat{\bar{\boldsymbol{\alpha}}}$ for the state $\psi_\zeta$, Eq.\ (\ref{Bell}).  We have therefore suppressed these indices of $ {\rho}^{{\rm  spin} }$. For the proposed  Bell test  one needs to measure the pseudospin correlators $C_{ab}$, Eq.\ (\ref{Cabdef}), evaluated for the postselected electron-hole pairs. They   follow from  
\beq \label{Cabuu}
C^{\uparrow}_{ ab}=\frac{1}{4}\,{\rm Tr}\,\rho^{\rm spin} (1+\hat{\boldsymbol{a}}\cdot \boldsymbol{\sigma^{\rm el}}  )(1+\hat{\boldsymbol{b}} \cdot \boldsymbol{\sigma^{\rm h}} )
\eeq
as $C_{ab}=C^{\uparrow}_{ ab}+C^{\uparrow}_{ ba}-C^{\uparrow}_{ -ab}-C^{\uparrow}_{ b\, -a}-C^{\uparrow}_{ a\,-b}-C^{\uparrow}_{-b a}+C^{\uparrow}_{ -a\,-b}+C^{\uparrow}_{ -b\,-a}$. We assume $\Omega\gg v/\Delta r \gg  {\rm max} \{\Omega(eV_{\rm ex}/\varepsilon_{\rm F})^2 ,kT(\varepsilon_{\rm F}/eV_{\rm ex})^2 (kT/\Omega)\exp(-2\omega/kT)  \}$. In this limit it is very unlikely for two holes to be within a distance $\Delta r$ of the same electron (or vice versa).  The correlators $C^{\uparrow}_{ ab}$ may  then be measured by ``coincidence detection'' \cite{kawabata:jps01,chtchelkatchev:prb02}, correlating momentum-projected electron and hole   pseudospin densities,
 \beq \label{dens0}
C^{\uparrow}_{ ab} \propto \int_{0}^R  d {r} d\bar{ {r}}   \,e^{-(r-\bar{r})^2/2\Delta r^2}  \, \langle n_{a}^{ \uparrow{\rm el} }( r) n_b^{\uparrow {\rm h}}(\bar{{r}}) \rangle.
 \eeq
Here,  the  densities $n^{ \uparrow{\rm el}}$ and  $n^{\uparrow {\rm h}}$  are defined in terms of the momentum-projected electron annihilation operators
 \beq \label{psi}
 \vec{\psi}_\alpha ^{\mu}( {r})=\int_0^\infty{ \frac{dp}{2\pi}\,e^{i pr}\,  \Theta[\gamma^\mu ( v {p}- \varepsilon_{\rm F})-\omega]\sum_{\eta=\pm   1 }   }  {\cal P}_{\eta\hat{\boldsymbol{\alpha}}}  \vec{\psi}_{ \eta p\hat{\boldsymbol{\alpha}} }
\eeq
as $n_\alpha ^{ \uparrow{ }\mu}(r)=[  \vec{\psi}_\alpha^{\mu\dag}(r)\cdot  \vec{v}^{\uparrow}_\alpha  ] [\vec{v}^{\uparrow\dag}_\alpha  \cdot \vec{\psi}_\alpha ^{\mu}(r) ]$. The index $\mu$ takes the values ${\rm el}$ or ${\rm h}$ and we have introduced $\gamma^{\rm el}=1$ and $\gamma^{\rm h}=-1$.  The vectors $v_\alpha ^\uparrow$  are the spinors corresponding to pseudospin up along the directions $\hat{\boldsymbol{\alpha}}$ and   the pseudospin matrix ${\cal P}_{\boldsymbol{p}} $  projects onto the conduction band. Note that at this point the choice of momentum direction  of the postselected excitations is arbitrary:  in the state (\ref{Bell}) amplitudes with any momentum have the same pseudospin. Our above choice of momenta along the desired pseudospin quantization axis $\hat{\boldsymbol{\alpha}}$ has been made merely for convenience. In our limit, when all electron-hole pairs are well-separated from each other (much farther than $\Delta r$), the density correlator of Eq.\ (\ref{dens0}) may be replaced by its irreducible contribution  
  \beq \label{densdel}
C^{\uparrow}_{ ab} \propto \int_{0}^R  dr d\bar{r} \,e^{-(r-\bar{r})^2/2\Delta r^2}  \, \langle \delta n_{a}^{ \uparrow{\rm el} }( r) \delta n_{b}^{\uparrow {\rm h}}(\bar{r}) \rangle
 \eeq  
\cite{kawabata:jps01,chtchelkatchev:prb02}. The  momentum-projected density $n_\alpha^{\mu}$ is  in principle experimentally accessible,  for instance by tunneling electrons and holes into additional reservoirs that couple  symmetrically to two contacts $\alpha$  and $-\alpha$ via momentum-dependent  tunnel amplitudes $w_\alpha^{\rm el}$ and $w_\alpha^{\rm h}$. Pseudospin-resolved  amplitudes $w^\mu$, as described by the tunneling Hamiltonian $H^{{\rm  el}}_{{\rm T}\alpha}+H^{{\rm  h}}_{{\rm T}\alpha}$ with
 \beq
 H^{\mu}_{{\rm T}\alpha}= \int{ dr \,    w_\alpha^\mu(r )\,   \vec{\psi}_\alpha^{\mu\dagger}(r)\cdot  \vec{\psi}^{{\rm res} \,\mu}_\alpha+h.c.},
 \eeq
then allow   to obtain the correlator $C^{\uparrow}_{ ab}$ as
  \beq \label{ccurr}
 C^{\uparrow}_{ ab} \propto c^{\uparrow{\rm el-h}}_{ab} + {\cal O}\left(e^{-\omega/kT}\right)
 \eeq  
 from current correlators
 \beq 
 c^{\uparrow{\rm el-h}}_{ab}= \int d t  \,e^{-(v t)^2/2\Delta r^2} \, \langle \delta I_{a}^{\uparrow{\rm el}}(t)\delta I_{b}^{\uparrow{\rm h}}(0) \rangle,
 \eeq
  where
   \beq
 I^{\uparrow\mu}_\alpha = i e\int{ dr \,    w_\alpha^\mu(r )\, [ \vec{\psi}_\alpha^{\mu\dagger}(r)\cdot  \vec{v}^\uparrow_\alpha  ]( \vec{v}^{\uparrow\dag}_\alpha \cdot  \vec{\psi}^{{\rm res} \,\mu }_\alpha)+h.c.}
 \eeq  
 We have introduced one reservoir  for every pseudospin state, with corresponding electron annihilation operators $\psi^{{\rm res}\, \mu }_{\alpha\sigma}$. The amplitudes  $w^\mu_\alpha$  are peaked at the radii $r^\mu_\alpha$. 
 In the step from Eq.\  (\ref{densdel}) to (\ref{ccurr}) we have assumed time-translational invariance and  $r^{\rm el}_\alpha=r^{\rm h}_\alpha$.
 Note that the error    due to thermal excitations flowing from the reservoirs into the graphene sheet (the origin of the problem pointed out in Ref.\ \cite{hannes:prb08}) is here exponentially suppressed. It is  of order $\exp(-\omega/kT)$ since it is only pairs of excitations   that differ in energy by at least $\omega$ that contribute to the correlator $c^{\uparrow{\rm el-h}}$.

 The correlator $C^{\uparrow}_{ ab}$, however, may be accessed also through measurements of  correlators
    \beq \label{cfinal}
 c^{\uparrow }_{ab}= \int d t  \,e^{-(v t)^2/2\Delta r^2} \, \langle \delta I_{a}^{\uparrow}(t)\delta I_{b}^{\uparrow}(0) \rangle
 \eeq
  of currents
   \beq
 I^\uparrow_\alpha = i e\int {dr \,  w_\alpha(r ) \, [ \vec{\psi}_\alpha ^{\dagger}(r)\cdot \vec{v}^\uparrow_\alpha ]( \vec{v}^{\uparrow\dag}_\alpha \cdot \vec{\psi}^{\rm res} _{\alpha}) +h.c.}
 \eeq
 without energy-selectivity, where  $ \vec{\psi}_\alpha (r)=\int_0^\infty{ \frac{dp}{2\pi} \, \exp(ipr) } \sum_{\eta=\pm   1 }  {\cal P}_{\eta\hat{\boldsymbol{\alpha}}}  \vec{\psi}_{ \eta p\hat{\boldsymbol{\alpha}} }$ (and a corresponding tunneling Hamiltonian $H^{}_{{\rm T}\alpha}$). This is seen easiest by  separating  two contributions to the  above current correlators from each other: First there are contributions due to the electron-hole pairs created by $V_{\rm ex}$.  We denote these contributions to $c^{\uparrow{\rm el-h}}$ and $c^{\uparrow}$  by a subscript ``ex'', $c^{\uparrow{\rm el-h}}_{\rm ex}$ and $c^{\uparrow }_{ \rm ex }$, respectively. Second, there are contributions from  statistical correlations  of excitations  due to the Pauli principle, denoted by a subscript ``stat,'' $c^{\uparrow{\rm el-h}}_{\rm stat}$ and $ c^{\uparrow }_{ \rm stat }$. We have $c^{\uparrow }=c^{\uparrow }_{ \rm ex }+c^{\uparrow }_{ \rm stat}$ and likewise  $c^{\uparrow{\rm  el-h} }=c^{\uparrow{\rm el-h} }_{ \rm ex }+c^{\uparrow{\rm el-h} }_{ \rm stat}$. The typical energy separation between the electrons and the holes created by $V_{\rm ex}$ is $  v p -  v \bar{p} \approx \Omega$. For the currents $I^{\uparrow}_{{\rm ex}\, \alpha}$ carried by these excitations   we thus may  approximate $I^\uparrow_{{\rm ex\, \alpha}}=I^{\uparrow{\rm el}}_{{\rm ex}\, \alpha}+I^{\uparrow{\rm  h}}_{{\rm ex}\, \alpha}+{\cal O}(\omega/\Omega)$ and  $ c^{\uparrow }_{ {\rm ex}\, ab}=c^{\uparrow{\rm el-h}}_{ {\rm ex}\,ab}+c^{\uparrow{\rm el-h}}_{ {\rm ex}\,ba} +{\cal O}(\omega/\Omega)$.   Under the assumption  $kT \gg v/|{r}^\mu_a \hat{\boldsymbol{a}} \pm {r}^\mu_b \hat{\boldsymbol{b}}|$ made in the main text  the statistical correlations $c^{\uparrow}_{\rm stat}$ are identical to  the equilibrium correlations, $c^{\uparrow}_{\rm stat}=c^{\uparrow}|_{V_{\rm ex}=0}+{\cal O}[kT \exp(-kT|{r}^\mu_a \hat{\boldsymbol{a}} \pm {r}^\mu_b \hat{\boldsymbol{b}}|/v)]$ (see Appendix B). 
 Moreover, statistical  fluctuations do not contribute to the momentum projected correlator by our definition of $ \vec{\psi}^{\mu}$: $\langle  {\psi}_{\alpha \sigma}^{{\rm el}\dag}(r,t) {\psi}^{{\rm h}}_{\alpha' \sigma'}(r',t')\rangle=0$.
 We conclude that to leading order in our limit the momentum projected irreducible current correlator $c^{\uparrow{\rm el-h} }$ may be expressed through the corresponding correlator without momentum projection after subtraction of its statistical background, $ c^{\uparrow{\rm el-h}}_{ab}+c^{\uparrow{\rm el-h}}_{ ba}   =c^{\uparrow{\rm el-h}}_{ {\rm ex}\,ab}+c^{\uparrow{\rm el-h}}_{ {\rm ex}\,ba}   =c^{\uparrow}_{{\rm ex}\,ab}=c^{\uparrow}_{ab}-c^{\uparrow}_{ab}\big|_{V_{\rm ex}=0}  $, such that
\beq \label{dens}
 C^{\uparrow}_{ ab}+C^{\uparrow}_{ ba} \propto  c^{ }_{ab}-c^{ }_{ab}\Big|_{V_{\rm ex}=0} .
 \eeq
 The  pseudospin-resolved tunneling amplitudes assumed above are rather unrealistic.  Alternatively the pseudospin  currents $I^{\uparrow}_\alpha $  can be measured by introducing separate reservoirs for the contacts  $\alpha$ and $-\alpha$, as explained after Eq.\ (\ref{project}) of the main text. One further shows straightforwardly along the lines of Refs.\ \cite{kawabata:jps01,chtchelkatchev:prb02} that the correlators $c^{ }$, Eq.\ (\ref{cfinal}) may be replaced by zero-frequency correlators [with an error  of  ${\cal O}(v/\Omega \Delta r)$]. Eqs.\ (\ref{Cab}) with (\ref{cex}) then follow after a normalization of $C^{\uparrow}$ along the lines of Refs.\ \cite{kawabata:jps01,chtchelkatchev:prb02}.   Values for $\omega$ and $\Delta r$ that satisfy all of the above conditions can be found provided that $kT \ll \Omega$. A violation of the inequality (\ref{CHSH}) with the    correlators Eqs.\ (\ref{Cab}) and (\ref{cex}) is thus indeed an entanglement witness to leading order in $ v/|{r}^\mu_a \hat{\boldsymbol{a}} \pm {r}^\mu_b \hat{\boldsymbol{b}}|\ll kT \ll \Omega$. 
 
 Strictly speaking, a Bell test has to demonstrate that a violation of  Eq.\ (\ref{CHSH}) found through the current correlation measurement described above is  due to the entanglement of $\rho^{\rm spin}$ rather than the deviations of the true pseudospin correlators from  Eq.\ (\ref{Cab}) that appear at nonzero temperature (even though those are suppressed in our limit). Collecting all the sources of such deviations mentioned above we find that the  Bell parameter ${\cal B}$ measured through Eq.\ (\ref{Cab}) is related to the Bell parameter ${\cal B}^{\rm spin}$ corresponding to the actual pseudospin correlators Eqs.\ (\ref{Cabdef}) and (\ref{Cabuu}) as
  \begin{widetext}
 \bea \label{Bac}
 {\cal B} &=&  {\cal B}^{\rm spin}(\omega,\Delta r) + g_1\,  \frac{\omega}{\Omega}+g_2 \,\frac{v}{\Omega\Delta r }+g_3\, \frac{\Omega\Delta r }{v}\left(\frac{eV_{\rm ex}}{\varepsilon_{\rm F}}\right)^2 +g_4\, \frac{kT \Delta r }{v} \frac{kT}{\Omega}\left(\frac{\varepsilon_{\rm F}}{eV_{\rm ex}}\right)^2 e^{-2\omega/kT} + \nonumber \\
 && + \sum_{\alpha \alpha'} g_{\alpha\alpha'} \,\frac{kT}{\Omega} \left(\frac{\varepsilon_{\rm F}}{eV_{\rm ex}}\right)^2 e^{-kT|r_\alpha \boldsymbol{\hat{\alpha}} - r_{\alpha'} \boldsymbol{\hat{\alpha}}'|/v}
 \eea
 up to terms of higher order in small quantities, with positive constants $g_j $  and $g_{\alpha\alpha'}$ that are of order unity. The summation in Eq.\ (\ref{Bac}) runs over all measured combinations of tunnel contacts $\alpha$, $\alpha'$. Entanglement   is conclusively demonstrated if one finds a violation of the inequality ${\cal B}^{\rm spin} \leq 2$, that is if
  \beq \label{CHSHac}
 {\cal B} \leq 2 + g_1\,  \frac{\omega}{\Omega}+g_2 \,\frac{v}{\Omega\Delta r }+g_3\, \frac{\Omega\Delta r }{v}\left(\frac{eV_{\rm ex}}{\varepsilon_{\rm F}}\right)^2 +g_4\, \frac{kT \Delta r }{v} \frac{kT}{\Omega}\left(\frac{\varepsilon_{\rm F}}{eV_{\rm ex}}\right)^2 e^{-2\omega/kT} + \sum_{\alpha \alpha'} g_{\alpha\alpha'} \,\frac{kT}{\Omega} \left(\frac{\varepsilon_{\rm F}}{eV_{\rm ex}}\right)^2 e^{-kT|r_\alpha \boldsymbol{\hat{\alpha}} - r_{\alpha'} \boldsymbol{\hat{\alpha}}'|/v}.
 \eeq
 \end{widetext}
 A direct violation of Eq.\ (\ref{CHSHac}) requires knowledge of the parameters $g_j$ and $g_{\alpha\alpha'}$. They   can in principle be determined if the frequency spectrum of the voltage correlator $c_V$ is known. More practically, however, one may measure the dependence of ${\cal B}$ on parameters such as $T$ and $V_{\rm ex}$  and establish an inconsistency with Eq.\ (\ref{CHSHac}) for a particular choice of $\omega$ and $\Delta r$.   
 For example, one may  choose   $\omega=\sqrt{kT\Omega}$ and $\Delta r= \varepsilon_{\rm F} v/\Omega eV_{\rm ex}$, such that 
  \begin{widetext}
 \bea \label{Bomr}
 {\cal B} &=&  {\cal B}^{\rm spin}(\sqrt{kT\Omega}, \varepsilon_{\rm F} v/\Omega eV_{\rm ex}) + g_1\,  \sqrt{\frac{kT}{\Omega}}+(g_2+g_3) \,\frac{eV_{\rm ex}}{\varepsilon_{\rm F}} +g_4\, \ \left(\frac{kT}{\Omega}\right)^2\left(\frac{\varepsilon_{\rm F}}{eV_{\rm ex}}\right)^3 e^{-2\sqrt{\Omega/kT}} + \nonumber \\
 && + \sum_{\alpha \alpha'} g_{\alpha\alpha'} \,\frac{kT}{\Omega} \left(\frac{\varepsilon_{\rm F}}{eV_{\rm ex}}\right)^2 e^{-kT|r_\alpha \boldsymbol{\hat{\alpha}} - r_{\alpha'} \boldsymbol{\hat{\alpha}}'|/v}.
\eea
An analysis of  ${\rho}^{{\rm  spin} }$, Eqs.\ (\ref{rho}) with (\ref{rhospin}), shows for the same choice of $\omega$ and $\Delta r$ 
\bea \label{B0}
  {\cal B}^{\rm spin}(\sqrt{kT\Omega}, \varepsilon_{\rm F} v/\Omega eV_{\rm ex})& =& {\cal B}^{\rm spin}(\sqrt{kT_0\Omega}, \varepsilon_{\rm F} v/\Omega eV_{0})  + \tilde{g}_1\, \left( \sqrt{\frac{kT}{\Omega}}-\sqrt{\frac{kT_0}{\Omega}}\right)+\tilde{g}_3 \,\left(\frac{eV_{\rm ex}}{\varepsilon_{\rm F}} -\frac{eV_{0}}{\varepsilon_{\rm F}} \right)\nonumber \\
 && \mbox{}+\tilde{g}_4\, \ \left[\left(\frac{kT}{\Omega}\right)^2\left(\frac{\varepsilon_{\rm F}}{eV_{\rm ex}}\right)^3 e^{-2\sqrt{\Omega/kT}} -\left(\frac{kT_0}{\Omega}\right)^2\left(\frac{\varepsilon_{\rm F}}{eV_{0}}\right)^3 e^{-2\sqrt{\Omega/kT_0}} 
\right]
\eea
\end{widetext}
 up to terms of higher order in small quantities, with positive constants $\tilde{g}_j $    that are of order unity.
 Suppose that with this choice of $\omega$ and $\Delta r$ one measures  a parameter ${\cal B}-2>0$ that varies only by a small fraction $\Delta {\cal B}/({\cal B}-2)\ll 1$ over a   range of temperatures $T\in[T_0,2T_0]$ and potentials $|V_{\rm ex}|\in[V_0,2V_0]$ ($V_0>0$). Eqs.\ (\ref{Bomr}) with (\ref{B0}) prove this measurement result to be inconsistent with the assumption that no entanglement is present in the electron-hole pairs measured at $T_0$ and $V_0$. It excludes that  ${\cal B}^{\rm spin}(\sqrt{kT_0\Omega}, \varepsilon_{\rm F} v/\Omega eV_{0}) \leq 2$ (note that according to Eqs.\ (\ref{Bomr}) and (\ref{B0}) all $T$- and $V_{\rm ex}$-independent contributions to ${\cal B}$ have a negative sign). The result ${\cal B}-2>0$ with variation $\Delta {\cal B}/({\cal B}-2)\ll 1$ over the above parameter range, however, is predicted to be found    in the presence of entanglement (for appropriate choices of the pseudospin projection directions such as shown in Fig.\ \ref{fig1}). In such a measurement  the generated entanglement can thus be verified conclusively.

 \section{Appendix B: Current correlators}
 
In this Appendix we obtain the correlators $\tilde{c}_{ab}=\int dt\, \langle \delta I_a(t)\delta I_b(0)\rangle$   from the Hamiltonian $H=H_0+H_{\rm ex}+\sum_\alpha  H_{{\rm T},\alpha}$. Lowest order perturbation theory in $\vec{w}$ and $V_{\rm ex}$ results in 
 \begin{widetext}
 \beq \label{cKel0}
  c^{(0)}_{ab}= -\frac{e^4}{2} \int d\boldsymbol{x}_a\,d\boldsymbol{x}'_a\,d\boldsymbol{x}_b\, d\boldsymbol{x}'_b\,  \frac{d\varepsilon}{2\pi}\frac{d\varepsilon'}{2\pi} \, {\rm Tr}\,   \check{g}(\boldsymbol{x}'_b-\boldsymbol{x}_{a},\varepsilon)      [\check{\tau}^z,\check{G}_a(\varepsilon)] \check{W}_a(\boldsymbol{x}_a,\boldsymbol{x}'_a)\check{g}(\boldsymbol{x}'_a-\boldsymbol{x}_{b},\varepsilon')     [\check{\tau}^z,\check{G}_b(\varepsilon)] \check{W}_b(\boldsymbol{x}_b,\boldsymbol{x}'_b), 
   \eeq
describing statistical correlations. Here, $\check{g}$ and $\check{G}_\alpha $, are the Green functions of electrons in the graphene sheet and in reservoir $\alpha$, respectively. They are matrices with Keldysh-  \cite{rammer:rmp86} and pseudospin indices.  We expand the Green functions $\check{g}(\boldsymbol{x})$ asymptotically assuming $k_{\rm F} |\boldsymbol{x}|\gg 1$ \cite{mariani:prb07}, as it is appropriate in our limit $k_{\rm F} {r}_\alpha \gg 1$.  The matrix $\check{\tau}^z$ is the third Pauli matrix in   Keldysh space and the unit matrix in pseudospin space, while $\check{W}_\alpha $ is unity in Keldysh space and 
 \beq
 \check{W}_\alpha (\boldsymbol{x}_\alpha ,\boldsymbol{x}_\alpha ')  = \left(\begin{array}{cc} w^{\rm A}_\alpha (\boldsymbol{x}_\alpha ) w^{{\rm A}*}_\alpha (\boldsymbol{x}_\alpha ') &  w^{\rm A}_\alpha (\boldsymbol{x}_\alpha ) w^{{\rm B}*}_\alpha (\boldsymbol{x}_\alpha ')  \\ w^{\rm B}_\alpha (\boldsymbol{x}_\alpha ) w^{{\rm A}*}_\alpha (\boldsymbol{x}_\alpha ') &  w^{\rm B}_\alpha (\boldsymbol{x}_\alpha ) w^{{\rm B}*}_\alpha (\boldsymbol{x}_\alpha ')\end{array}\right)
 \eeq
 in pseudospin space. 
 
 At the next to leading (second) order in $V_{\rm ex}$ we find a contribution
\bea \label{cKela}
 && c^{(2){\rm ex}}_{ab}= -\frac{e^4}{2} \int d\boldsymbol{x}_a\,d\boldsymbol{x}'_a\,d\boldsymbol{x}_b\, d\boldsymbol{x}'_b\, d \boldsymbol{x}_{\rm ex}\,d\boldsymbol{x}'_{\rm ex}\, \frac{d\varepsilon}{2\pi}\frac{d\varepsilon'}{2\pi} \, c_{V}(\varepsilon-\varepsilon')u(\boldsymbol{x}_{\rm ex}) u(\boldsymbol{x}'_{\rm ex})
\, \\
 &&\;\;\mbox{} \times {\rm Tr}\,   \check{g}(\boldsymbol{x}'_b-\boldsymbol{x}_{\rm ex},\varepsilon)    \check{\tau}^z \check{g}(\boldsymbol{x}_{\rm ex}-\boldsymbol{x}_a,\varepsilon') [\check{\tau}^z,\check{G}_a(\varepsilon')] \check{W}_a(\boldsymbol{x}_a,\boldsymbol{x}'_a)\check{g}(\boldsymbol{x}'_a-\boldsymbol{x}'_{\rm ex},\varepsilon')   \check{\tau}^z \check{g}(\boldsymbol{x}'_{\rm ex}-\boldsymbol{x}_b,\varepsilon) [\check{\tau}^z,\check{G}_b(\varepsilon)] \check{W}_b(\boldsymbol{x}_b,\boldsymbol{x}'_b)\nonumber
   \eea
 due to electron-hole pairs excited by the fluctuating potential $V_{\rm ex}$ and a contribution
 \bea \label{cKelstat}
 && c^{(2){\rm stat}}_{ab}= -\frac{e^4}{2} \int d\boldsymbol{x}_a\,d\boldsymbol{x}'_a\,d\boldsymbol{x}_b\, d\boldsymbol{x}'_b\, d \boldsymbol{x}_{\rm ex}\,d\boldsymbol{x}'_{\rm ex}\, \frac{d\varepsilon}{2\pi}\frac{d\varepsilon'}{2\pi} \, c_{V}(\varepsilon-\varepsilon')u(\boldsymbol{x}_{\rm ex}) u(\boldsymbol{x}'_{\rm ex})
\, \\
 &&\; \mbox{} \times {\rm Tr}\,  \left\{ \check{g}(\boldsymbol{x}'_b-\boldsymbol{x}_{\rm ex},\varepsilon)    \check{\tau}^z \check{g}(\boldsymbol{x}_{\rm ex}-\boldsymbol{x}_{\rm ex}',\varepsilon')  \check{\tau}^z \check{g}(\boldsymbol{x}'_{\rm ex}-\boldsymbol{x}_a,\varepsilon)  [\check{\tau}^z,\check{G}_a(\varepsilon)] \check{W}_a(\boldsymbol{x}_a,\boldsymbol{x}'_a)\check{g}(\boldsymbol{x}'_a-\boldsymbol{x}_{b},\varepsilon) [\check{\tau}^z,\check{G}_b(\varepsilon)] \check{W}_b(\boldsymbol{x}_b,\boldsymbol{x}'_b)\right.\nonumber \\
&& \;\;\;\;\;\;\;\mbox{} +\left.  \check{g}(\boldsymbol{x}'_b-\boldsymbol{x}_{a},\varepsilon) [\check{\tau}^z,\check{G}_a(\varepsilon)] \check{W}_a(\boldsymbol{x}_a,\boldsymbol{x}'_a)\check{g}(\boldsymbol{x}'_a-\boldsymbol{x}_{\rm ex},\varepsilon)    \check{\tau}^z \check{g}(\boldsymbol{x}_{\rm ex}-\boldsymbol{x}_{\rm ex}',\varepsilon')  \check{\tau}^z \check{g}(\boldsymbol{x}'_{\rm ex}-\boldsymbol{x}_b,\varepsilon) [\check{\tau}^z,\check{G}_b(\varepsilon)] \check{W}_b(\boldsymbol{x}_b,\boldsymbol{x}'_b)\nonumber\right\}
   \eea
   \end{widetext}
which describes statistical correlations due to a renormalization of the tunneling amplitudes $\vec{w}$ by $V_{\rm ex}$.
 
     To lowest order in our limit  $\Omega\ll \varepsilon_{\rm F}$ and  $l_{\rm ex}, l_\alpha  \ll |\boldsymbol{r}_\alpha |$ we have $w^{\sigma}_\alpha (\boldsymbol{x}   ) =(v/\sqrt{k_{\rm F}}) \bar{w}^{\sigma}_\alpha  \delta(\boldsymbol{x}   - r_\alpha \boldsymbol{\hat{\alpha}} )$ and $u(\boldsymbol{x})= \delta(\boldsymbol{x})/k^2_{\rm F}$.  Eq.\ (\ref{cKela}) then evaluates to
\beq \label{cresapp}
\tilde{c}^{(2){\rm ex}}_{{ a},{ b}}= \frac{e^4}{2\varepsilon_{\rm F}^2} (1+\hat{\boldsymbol{a}}\cdot \hat{\boldsymbol{b}}) W_{ a}W_{ b} \int_{-\infty}^{\varepsilon_{\rm F}}d\varepsilon \int_{\varepsilon_{\rm F}}^{\infty}d\varepsilon' \,c_V(\varepsilon-\varepsilon')
 \eeq
 with the tunneling probabilities
\beq
W_{ \alpha} =  \varepsilon_{\rm F}\frac{ \left|\bar{w}^{\rm A}_\alpha \right|^2 +  \left|\bar{w}^{\rm B}_\alpha \right|^2 - 2  \left|\bar{w}^{\rm A}_\alpha  \bar{w}^{\rm B}_\alpha \right| (\cos \nu_\alpha ,\sin \nu_\alpha )\cdot \hat{\boldsymbol{ \alpha}} }{4\pi^2 v_\alpha  k^2_{\rm F}  r_\alpha}  ,
\eeq
where $\nu_\alpha  =i  \ln\left( \bar{w}^{{\rm A}}_\alpha \bar{w}^{{\rm B}*}_\alpha /\left|\bar{w}^{\rm A}_\alpha  \bar{w}^{\rm B}_\alpha \right|\right)$ and $v_\alpha $ is the Fermi velocity in reservoir $\alpha$.  
 The above perturbation expansion in $\vec{w}$ is justified if $\bar{w} \ll 1$.

 The contribution $c^{(2){\rm stat}}$, Eq.\ (\ref{cKelstat}), is exponentially suppressed as  $\exp(-kT |r_a \boldsymbol{\hat{a}} \pm r_b \boldsymbol{\hat{b}}|/v)$ by thermal dephasing in the limit $kT \gg v/|r_a \boldsymbol{\hat{a}} \pm r_b \boldsymbol{\hat{b}}|$ taken in the main text. After the subtraction of the statistical fluctuations $c^{(0) }$ and the normalization performed in Eq.\ (\ref{cex}) the correlator $c_{a,b}$  takes the form of Eq.\ (\ref{cres}) to this accuracy. Similarly we obtain for the AC-current into  tunnel contact $\alpha$ 
\begin{widetext}
\beq
I_\alpha (\omega)=ie^2 V_{\rm ex}(\omega)  \int d\boldsymbol{x}_\alpha \,d\boldsymbol{x}'_\alpha \,d\boldsymbol{x}_{\rm ex}\,  \frac{d\varepsilon}{2\pi}  \,u(\boldsymbol{x}_{\rm ex})\,  {\rm Tr}\,   \check{g}(\boldsymbol{x}'_\alpha -\boldsymbol{x}_{\rm ex},\varepsilon)   \check{\tau}^z   \check{g}(\boldsymbol{x}_{\rm ex}-\boldsymbol{x}_{\alpha},\varepsilon-\omega)  [\check{\tau}^z\check{G}_\alpha (\varepsilon)-\check{G}_\alpha (\varepsilon-\omega)\check{\tau}^z] \check{W}_\alpha (\boldsymbol{x}_\alpha ,\boldsymbol{x}'_\alpha ), 
   \eeq
\end{widetext}
which implies Eq.\ (\ref{W}) in our limit.
 \end{appendix}  

\begin{thebibliography}{28}
\bibitem[{foo({\natexlab{a}})}]{footnote0}
\bibinfo{note}{Purely electrical implementations of fluctuating potentials in
  the frequency range $\Omega \gg kT$ are challenging, but have been achieved
  in the past \cite{schoelkopf:prl98,reydellet:prl03}. Alternatively one could
  imagine to generate the potential $V_{\rm ex}$ by a tilted near-field tip or
  by a metallic nanoparticle at a temperature $T'\gg T$ in the vicinity of the
  graphene sheet (possibly partially shielded to maintain $\langle (eV_{\rm
  ex})^2\rangle\ll \varepsilon_{\rm F}^2$).}
\bibitem[{foo({\natexlab{b}})}]{footnote01}
\bibinfo{note}{The single-valley form of $H_{{\rm T},j}$ exploits the time
  reversal invariance of the tunneling Hamiltonian (at $B=0$). The spinor
  $\vec{w}_j$ is obtained from the corresponding spinor $(\tilde{w}^{\rm A}_j
  ,\tilde{w}^{\rm B}_j ,\tilde{w}'^{\rm A}_j ,\tilde{w}'^{\rm B}_j)$ in the
  original two-valley model by $ {w}^{\rm A}_j=[\tilde{w}^{\rm
  A}_j+(\tilde{w}'^{\rm A}_j)^*]/\sqrt{2}$ and $ {w}^{\rm B}_j=[\tilde{w}^{\rm
  B}_j-(\tilde{w}'^{\rm B}_j)^*]/\sqrt{2}$.}
\bibitem[{foo({\natexlab{c}})}]{footnote2}
\bibinfo{note}{A rigorous quantum mechanical treatment of the problem that
  projects the Bell pair wavefunction Eq.\ (\ref{Bell}) onto the eigenstates of
  $\boldsymbol{x}(t)$, Eq.\ (\ref{project}), to leading order in $(k_{\rm
  F}{r}_j)^{-1}$ confirms this semiclassical reasoning. In momentum space and
  polar coordinates the eigenstates to eigenvalue $\boldsymbol{r}=(r,\phi)$
  take the form $\psi_\pm( p,\theta)=e^{\pm ivtp-i pr \cos(\phi-\theta)} (1,\pm
  e^{i \theta})$ to this accuracy.}
\bibitem[{foo({\natexlab{d}})}]{footnote21}
\bibinfo{note}{We have disregarded the electron spin in our discussion. It
  enters Eqs.\ (\ref{W}) and (\ref{cres}) through numerical factors.}
\bibitem[{foo({\natexlab{e}})}]{footnote3}
\bibinfo{note}{A violation of a Bell inequality in the proposed experiment
  therefore does not disprove ``local realism.'' It is only used as a signature
  of the produced entanglement.}
\end{thebibliography}
   \end{document}